\pdfoutput=1
\documentclass{article}

\usepackage{arxiv}

\usepackage[utf8]{inputenc}
\usepackage[T1]{fontenc}
\usepackage{amsmath}
\usepackage{amssymb}
\usepackage{graphicx}
\usepackage{xcolor}
\usepackage{hyperref}
\usepackage{url}
\usepackage{cite}

\title{Inverse Design of Multi-Layer Sub-Pixel-Resolution RF Passives
Through Grayscale Diffusion with Flexible S-Parameter Conditioning}

\renewcommand{\shorttitle}{Inverse Design of RF Passives via Grayscale Diffusion}

\author{\normalfont
  Tommaso Dreossi$^{*}$, Christopher M. Bryant$^{*}$, Hao Liu, Nathan Mirman,\\
  Noah Kessler, Michael Frei, Harish Krishnaswamy \\
  Arena Physica\\
  New York, NY, USA \\[4pt]
  {\small $^{*}$Equal contribution.}
}

\hypersetup{
pdftitle={Inverse Design of Multi-Layer Sub-Pixel-Resolution RF Passives Through Grayscale Diffusion with Flexible S-Parameter Conditioning},
pdfauthor={Tommaso Dreossi, Christopher M. Bryant, Hao Liu, Nathan Mirman, Noah Kessler, Michael Frei, Harish Krishnaswamy},
pdfkeywords={diffusion models, generative AI, inverse design, microwave filters, printed circuits, scattering parameters},
}

\begin{document}
\maketitle

\begin{abstract}
Inverse design of RF passive components from S-parameters is a high-dimensional, ill-posed problem, and prior generative approaches are limited to single-layer binary-metallization structures. This
paper presents an inverse design approach that generates passive components from partial S-parameter inputs on an 8$\times$8 mm
board discretized at 64$\times$64 pixels with sub-pixel grayscale metallization across 1--20 GHz. The framework generates
two-layer copper layouts with vias, with hard physical constraints on feed locations enforced through annealed Langevin
projection, flexible multi-modal conditioning on partial S-parameter specifications, port locations, dielectric
properties, reference topology, and variable port placement. Candidate designs are generated in seconds, with a Vision Transformer (ViT)-based forward model predicted S-parameters matching targets to within 0.77 $\pm$ 1.28 dB weighted mean absolute error. We validate the approach with two fabricated designs on RO4003C: a manufacturable alternative to a hairpin filter
whose coupling gaps violate fabrication rules, and a combline bandpass filter designed from scratch given only target
S-parameters.
\end{abstract}


\section{Introduction}

The design of RF passive components is fundamental to modern wireless systems. Conventional design relies on iterative
electromagnetic (EM) simulation and expert-driven parameter tuning, with
design time scaling steeply as component complexity increases. Topology
optimization offers algorithmic automation, but requires full-wave
simulations, with single designs taking days to
converge~\cite{aage2017,jensen2011,molesky2018}.

\begin{figure}[h!]
    \centering
    \includegraphics[width=\linewidth]{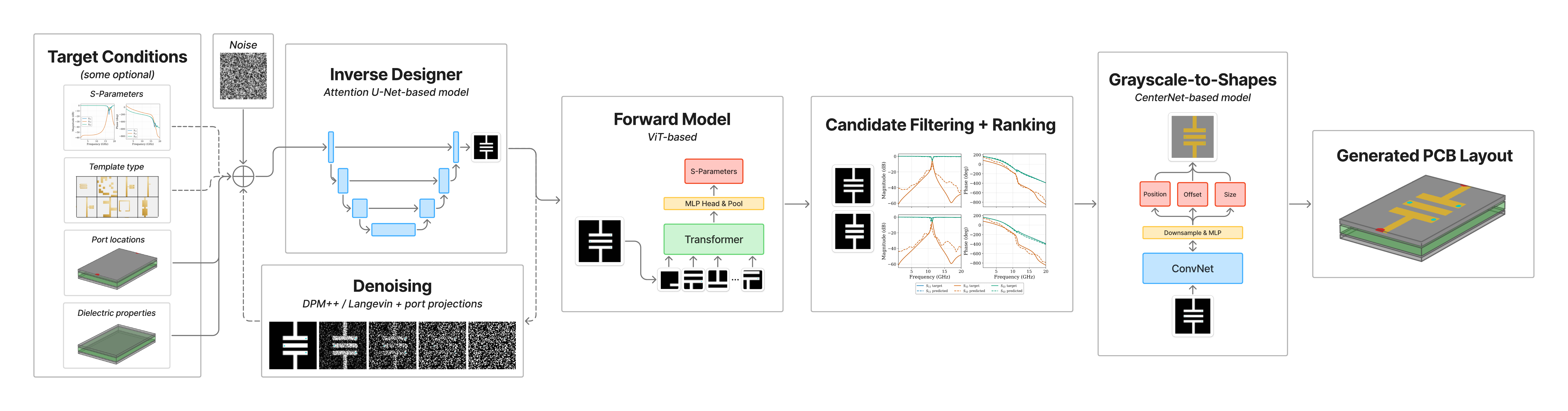} \\[-10pt]
    \caption{\textbf{Inverse design framework overview.} Target S-parameters, reference template type, port
locations, and dielectric properties are encoded and
concatenated channel-wise with the noisy board mask. The Attention U-Net iteratively denoises via DPM-Solver++ (or annealed Langevin dynamics with constraint projection) to produce candidate layouts, each comprising both a grayscale metallization and a vias layer (depicted in cyan). A Vision Transformer-based (ViT) forward model predicts the S-parameters of each candidate, which are ranked by agreement with the target S-parameters. A CenterNet-based model converts the rasterized metal layer into parameterized rectangle shapes exportable to common formats (e.g., Gerber).}
    \label{fig:system}
\end{figure}

Pre-computed Green's function methods accelerate individual
iterations~\cite{sun2026} but still require iterative optimization and
cannot produce diversity in design topology. Machine learning has emerged as a promising alternative. Neural Network-based surrogate or foundation models can potentially predict S-parameters from layout geometry orders of magnitude faster than full-wave simulation~\cite{karahan2023}, enabling inverse design via genetic algorithms~\cite{karahan2024}. More recently, generative models have been applied to directly synthesize
layouts from S-parameter specifications: Conditional Generative Adversarial Networks (GANs) produce microstrip
filter patterns from target responses~\cite{zhang2024,liu2018}, while the
DALL-EM framework introduced diffusion-based generation with a spatial entropy
design knob controlling pattern regularity~\cite{guo2025}. Separately, optimization-based approaches have addressed multi-layer on-chip passives
using direct binary search and particle swarm optimization, achieving
significant size reductions~\cite{chenna2025}, but remain bottlenecked by repeated full-wave EM simulations.

Despite this progress, all existing generative inverse design methods share fundamental limitations precluding practical PCB design. They operate on single-layer structures without vias, use binary (metal/no-metal) pixel representations with resolution limited to the grid size, are restricted to electrically small boards, require fixed port configurations, demand fully specified S-parameter targets, and ignore variable dielectric properties.

This paper presents a conditional diffusion framework (Fig.~\ref{fig:system}) that, for the first time in generative RF design, simultaneously supports:
\begin{itemize}
\item \textbf{Via-inclusive generation} of two-layer stackups.
\item \textbf{Grayscale continuous metallization} capturing sub-pixel features, critical for tightly coupled structures and complex signal paths.
\item \textbf{Variable port placement with hard constraints} enforced through annealed Langevin projection, guaranteeing physical realizability at feed points.
\item \textbf{Flexible multi-modal conditioning} on S-parameters, port locations, reference topology, and dielectric properties, with S-parameter and template inputs independently optional.
\end{itemize}

We validate the approach with two fabricated designs on RO4003C: a manufacturable alternative to a hairpin filter whose coupling gaps violate fabrication rules, and a combline bandpass filter designed from scratch given only target S-parameters.

\section{Inverse Design Framework}

\subsection{System Overview}

The proposed system comprises two neural networks operating in a
generate-and-rank pipeline (Fig.~\ref{fig:system}). A conditional denoising
diffusion probabilistic model (DDPM~\cite{ho2020,dhariwal2021}) generates candidate layouts
conditioned on target S-parameters, feed locations, dielectric properties,
and a template.
To select the best designs, a ViT-based forward model predicts the S-parameters of each candidate, compares them against the target
specification, and ranks them by root-mean-square error (RMSE) between
predicted and target responses.

The diffusion model is built on a U-Net architecture that outputs two channels:
metallization density and via presence. Unlike cross-attention conditioning
approaches~\cite{rombach2022}, all conditioning signals are encoded into
spatial feature maps and concatenated along the channel dimension with the
noisy layout, providing direct pixel-level conditioning while maintaining
computational efficiency. For unconstrained generation,
DPM-Solver++~\cite{lu2022} sampler produces candidates in seconds. When hard feed-point
constraints are required, the framework switches to annealed Langevin dynamics
with per-step projection (Section~\ref{sec:hardconstraint}).

\subsection{Training Data}

The model is trained on a dataset of approximately 3.5 million parametric PCB
layouts (over 24$\times$ larger than the 145K-sample dataset used by DALL-EM~\cite{guo2025}, the largest prior generative EM design dataset) simulated with the MATLAB RF Toolbox full-wave solver at 51 frequency
points across 1--20\,GHz. The dataset spans 25 component families including
filters, matching networks, antennas, transmission line elements, and random
rectangular geometries.

Each reference topology (hereafter \emph{template}) has randomized geometric parameters within physically meaningful ranges. These designs are rasterized onto a $64\times64$ grid (0.125\,mm per pixel), where partial
pixel coverage produces sub-pixel geometric detail. Variable dielectric substrates (FR4,
RO4003C, and others) and feed locations are sampled during
generation. At training time, the dataset is further augmented with S-parameter-preserving transformations including rotations, reflections, port swapping, and the addition of electrically isolated random structures.

\subsection{Multi-Modal Conditioning}

The framework accepts four conditioning inputs (``Target Conditions'' in Fig.~\ref{fig:system}), each processed by a dedicated
encoder.

\textit{S-Parameter Conditioning} to specify target frequency response: Target
S-parameters are provided at 51 frequency points across
1--20\,GHz. An MLP encoder
processes the magnitude (dB, z-score normalized) and phase ($\sin/\cos$) of
each component jointly. A binary mask indicates which components contain valid
data, natively supporting 1-port and 2-port configurations.
S-parameter components are optionally masked during training,
enabling partial specifications at inference (e.g., constraining only $|S_{21}|$).

\textit{Template Conditioning} to guide structural topology: each training sample belongs to one of 25 parametric component families (e.g., hairpin filter, coupled-line bandpass, matching network). A learnable null embedding enables template-free generation, where the model discovers a suitable topology on its own.

\textit{Feed Location Conditioning} to enable variable port placement. Port
positions are specified as physical coordinates and encoded via a
per-port MLP. A feed mask supports variable port count
(1 or 2 active ports).

\textit{Dielectric Conditioning} to support different substrate materials: Substrate properties ($\varepsilon_r$, $\tan\delta$, thickness) are encoded via linear projection, supporting multiple
substrate materials (FR4, RO4003C, etc.).

\subsection{Grayscale and Via Support}

The diffusion model (``Inverse Designer'' in Fig.~\ref{fig:system}) outputs continuous pixel values in $[0,\,1]$,
representing metallization density. These grayscale values encode sub-pixel
geometric detail; for example, a pixel value of 0.3 can represent a metal edge
located 30\% of the way through that pixel, capturing features substantially finer than the
0.125\,mm grid resolution. This is
critical for coupled-line structures such as hairpin resonator filters, where
coupling strength depends sensitively on gap widths that may be smaller than
a single pixel. Prior generative
approaches~\cite{zhang2024,liu2018,guo2025} produce strictly binary patterns,
limiting their ability to represent tightly coupled structures. A CenterNet-based detection network~\cite{zhou2019} with differentiable rasterizer refinement converts grayscale images into axis-aligned rectangular primitives for fabrication (``Grayscale-to-Shapes'' in Fig.~\ref{fig:system}).

In addition to the copper metallization layer, the diffusion model outputs a
dedicated via channel that indicates via placements across the board.
A subset of training templates include vias with 50\% probability,
covering via-connected low-pass filters,
stub filters, and matching networks. To our knowledge, this is the first
generative ML system to handle via-inclusive multi-layer structures.

\subsection{Hard Constraint Projection}
\label{sec:hardconstraint}

Physical realizability requires that every generated layout has metallization
at all specified feed points, the sole hard physical constraint enforced by
the framework. When feed connectivity must be guaranteed, the framework switches to annealed Langevin dynamics~\cite{christopher2024} with per-step projection that clamps metallization at feed locations, ensuring generated designs satisfy the constraint (Fig.~\ref{fig:system}).

\section{Results}

We present results in four parts.  First, we demonstrate generation diversity across structurally distinct designs. Second, we quantitatively evaluate two sampling strategies, showing that constrained Langevin dynamics achieves 89\% valid board rate with sub-1\,dB WMAE. Third, we validate the end-to-end pipeline with two fabricated designs on RO4003C: a hairpin filter re-design that circumvents sub-resolution coupling gaps while preserving the target passband at 17\,GHz, and a combline bandpass filter with vias designed from scratch given only target S-parameters, achieving a 9.5\,GHz passband with 4.6\,dB insertion loss.

\subsection{Evaluation}

\begin{figure}[t]
  \centering
  \setlength{\tabcolsep}{2pt}
  \begin{tabular}{ccc}
  \scriptsize Generated PCB Layout & \scriptsize Magnitude $| \mathbf{S}^* |$ & \scriptsize Phase $\angle \mathbf{S}^*$ \\[1pt]
  \includegraphics[width=0.20\linewidth]{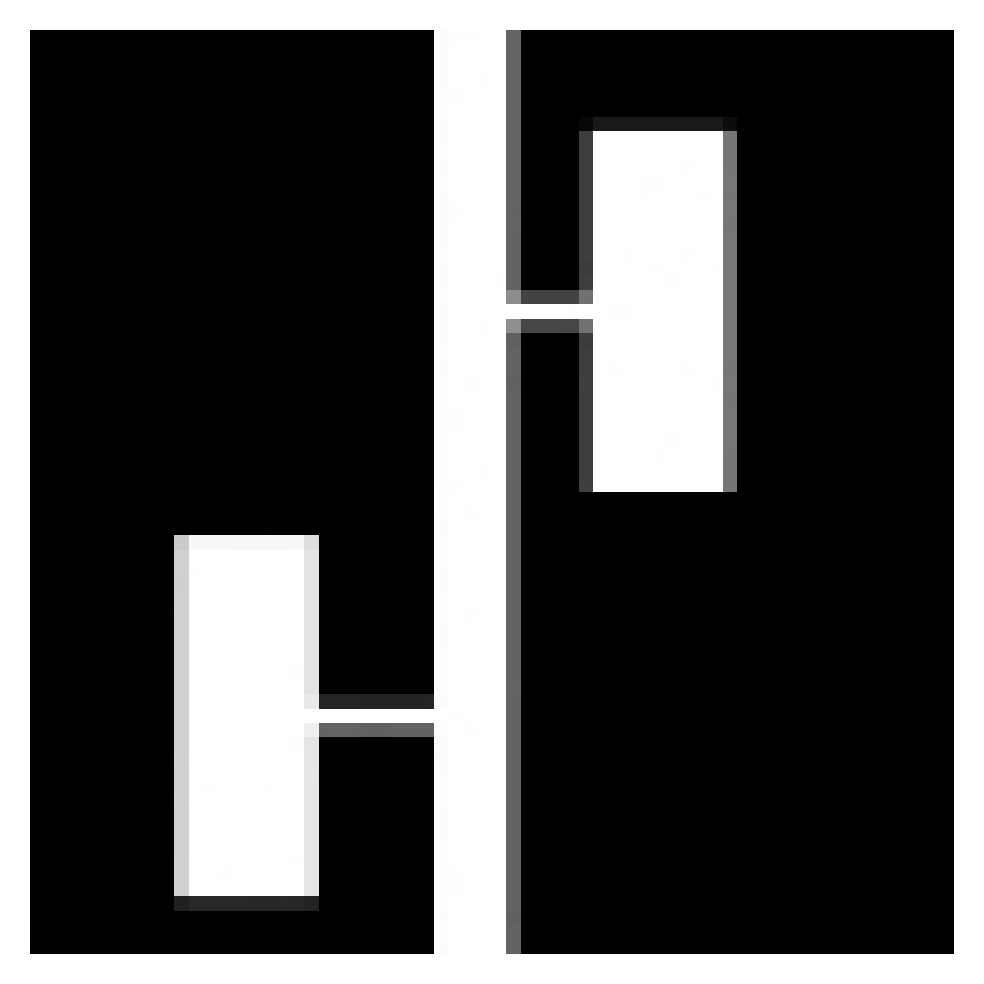} &
  \includegraphics[width=0.28\linewidth]{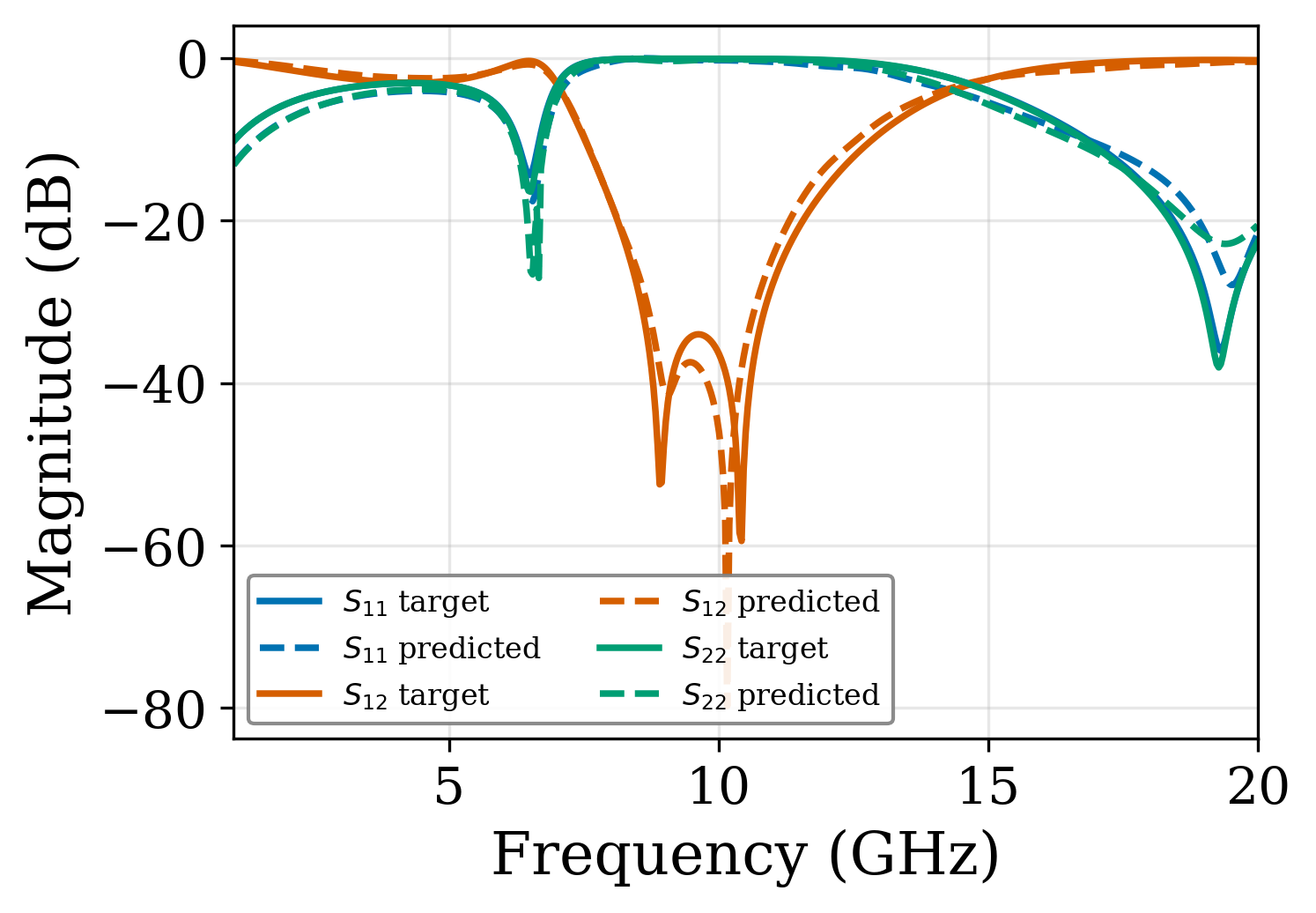} &
  \includegraphics[width=0.28\linewidth]{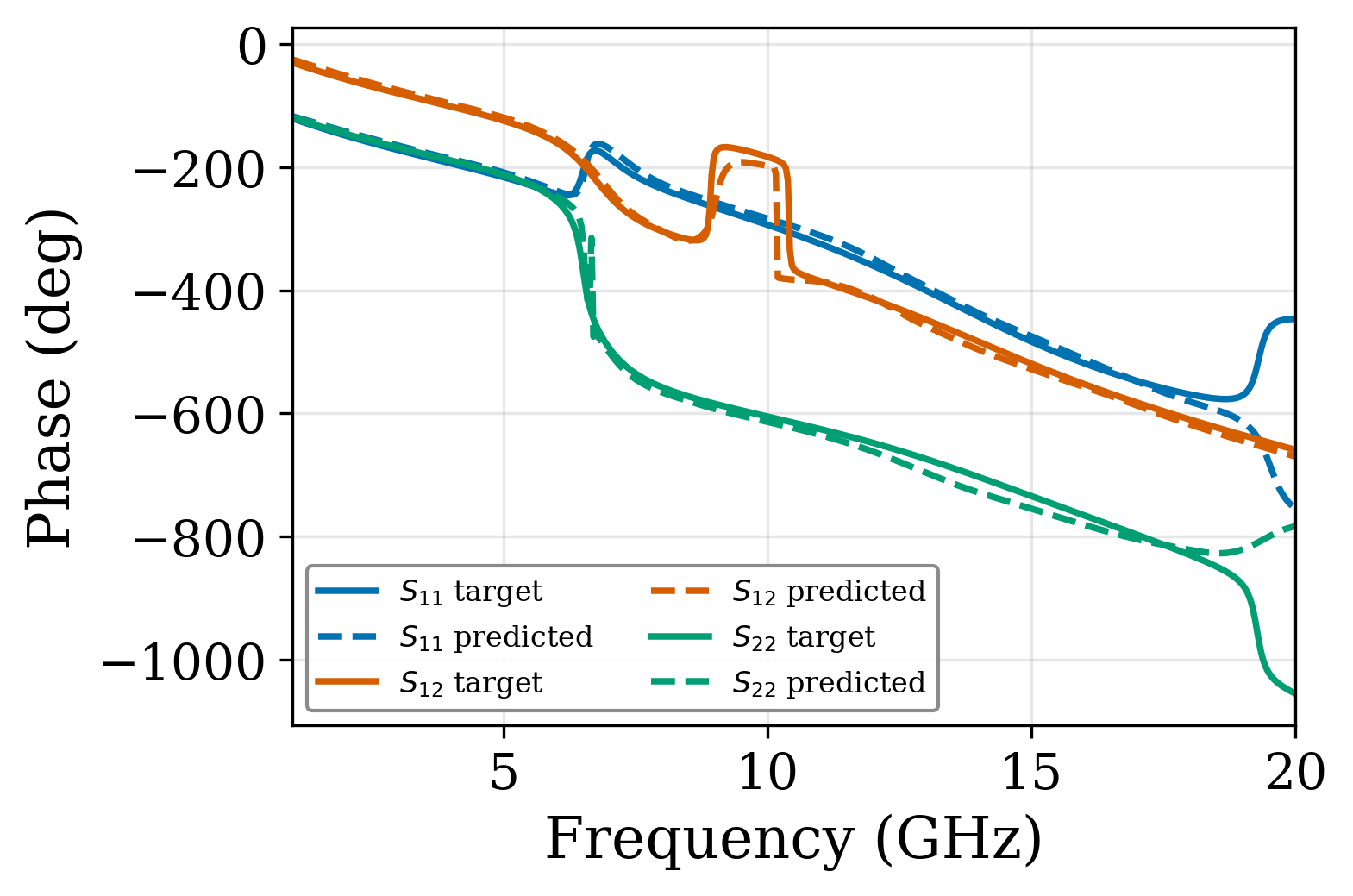}
  \\[0pt]
  \includegraphics[width=0.20\linewidth]{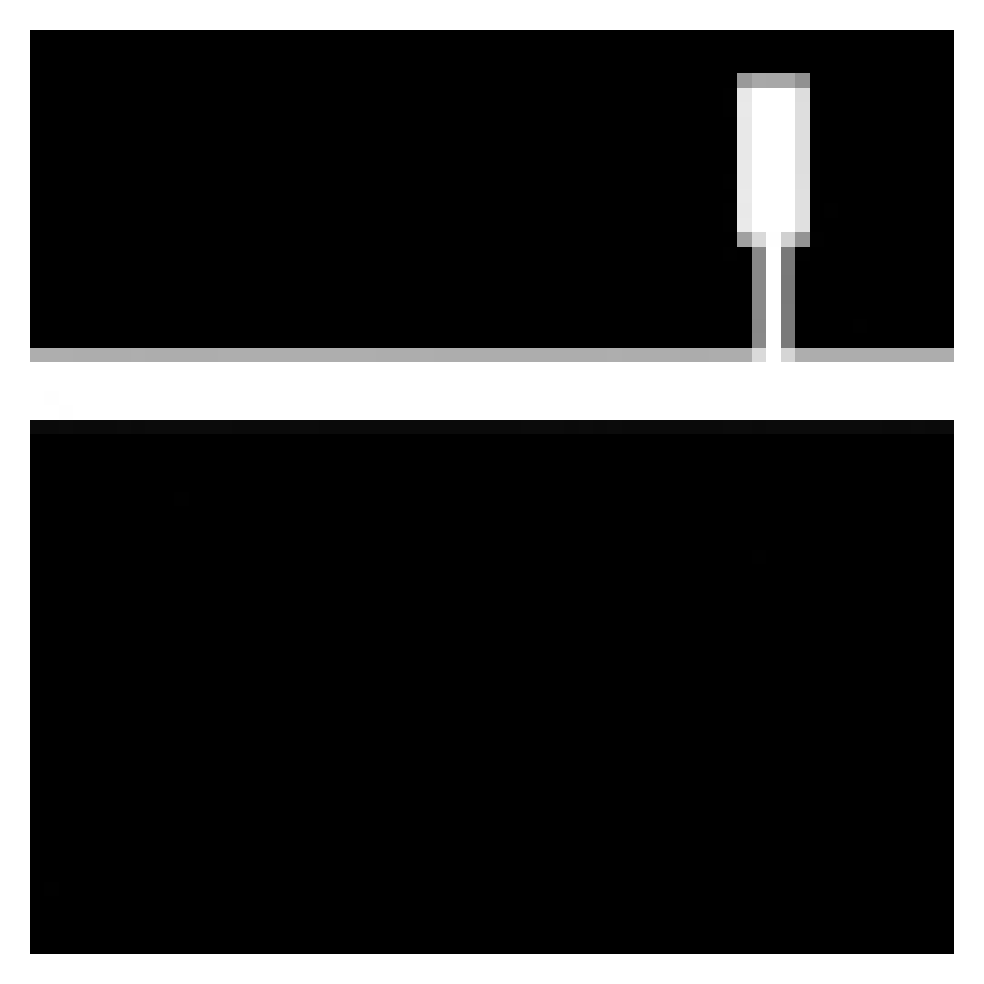} &
  \includegraphics[width=0.28\linewidth]{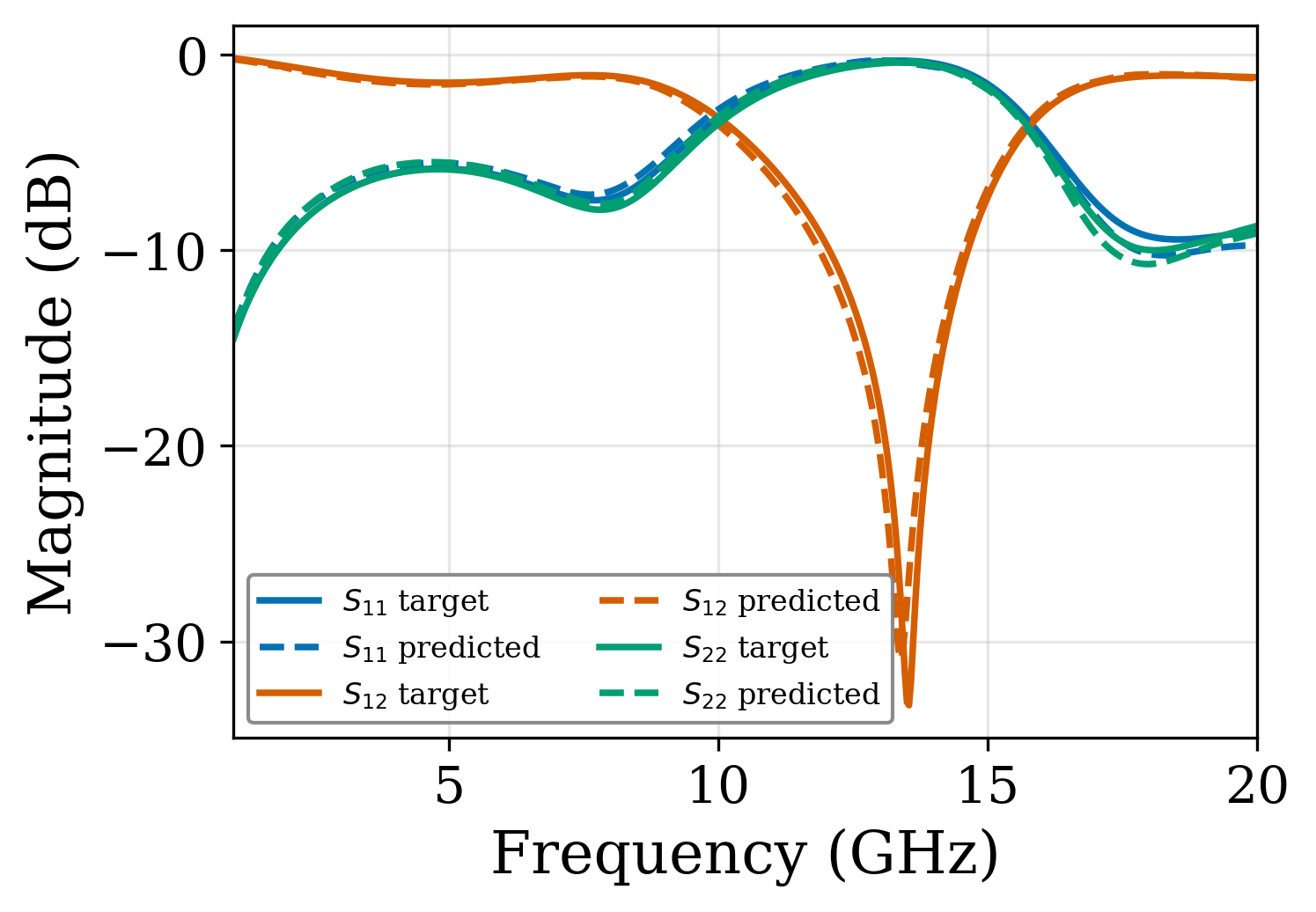} &
  \includegraphics[width=0.28\linewidth]{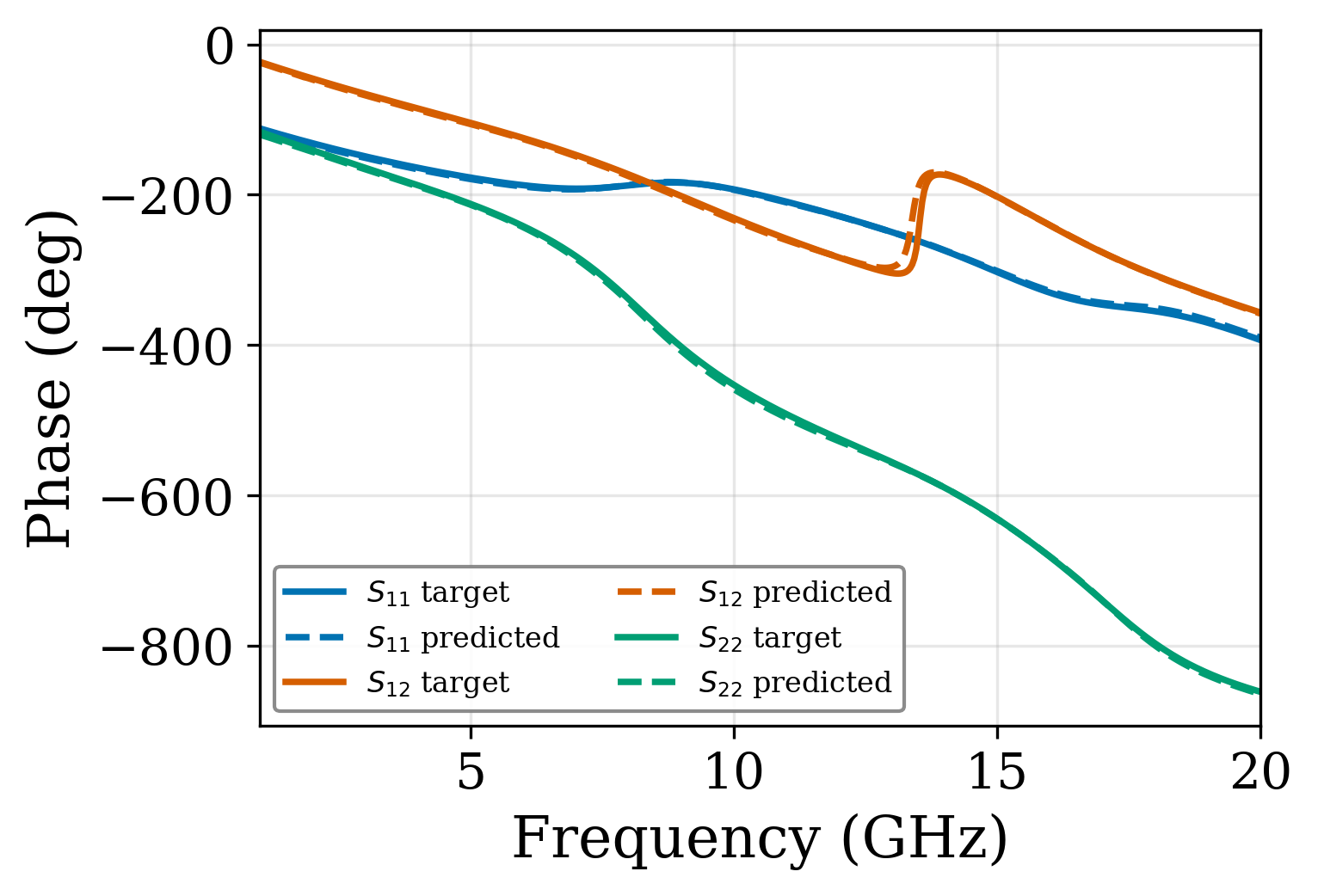}
  \\[1pt]
  \end{tabular}
  \caption{Examples of generated designs with
  S-Parameters predicted via the ViT-based forward model (dashed) vs.\ targets (solid).
  Columns show the generated PCB layout, S-parameter magnitude, and
  phase, respectively. Examples include a capacitive stub bandstop filter and an L-matching network with notch behavior.}
  \label{fig:gallery}
\end{figure}

We evaluated the generator on 50 validation samples using DPM++ (20 steps, 128 candidates) and annealed Langevin (1000 steps, 16 candidates).  Fig.~\ref{fig:gallery} shows representative generated designs with S-Parameters predicted via the ViT-based forward model overlaid on targets. Quality is measured by RMSE over real/imaginary S-parameter components and a weighted magnitude MAE (WMAE) in dB that de-emphasizes errors below $-20$\,dB.

Langevin sampling achieves a higher valid board rate (89.1\% vs.\ 8.5\%) due to its feed-projection constraints, which ensure metal coverage at port locations. Despite generating 8$\times$ fewer candidates, Langevin produces comparable best-candidate quality in terms of both forward model RMSE (0.149 $\pm$ 0.183 vs.\ 0.134 $\pm$ 0.185) and WMAE (0.93 $\pm$ 1.38\,dB vs.\ 0.77 $\pm$ 1.28\,dB). DPM++ is approximately 7$\times$ faster per sample (1.9s vs.\ 13.8s). These results suggest that Langevin is the preferred sampler for reliable inverse design, while DPM++ may be suitable for rapid exploration when validity can be enforced through post-processing.

\subsection{Design-for-Manufacturability: Hairpin Filter}

\begin{figure}[t]
\centering
    \setlength{\tabcolsep}{0pt}
    \begin{tabular}{@{}c@{}c@{}c@{}c@{}}
      Original & Generated & Manufactured & Magnitude $| \mathbf{S} |$ \\
      \raisebox{.2\height}{\includegraphics[width=.2\linewidth]{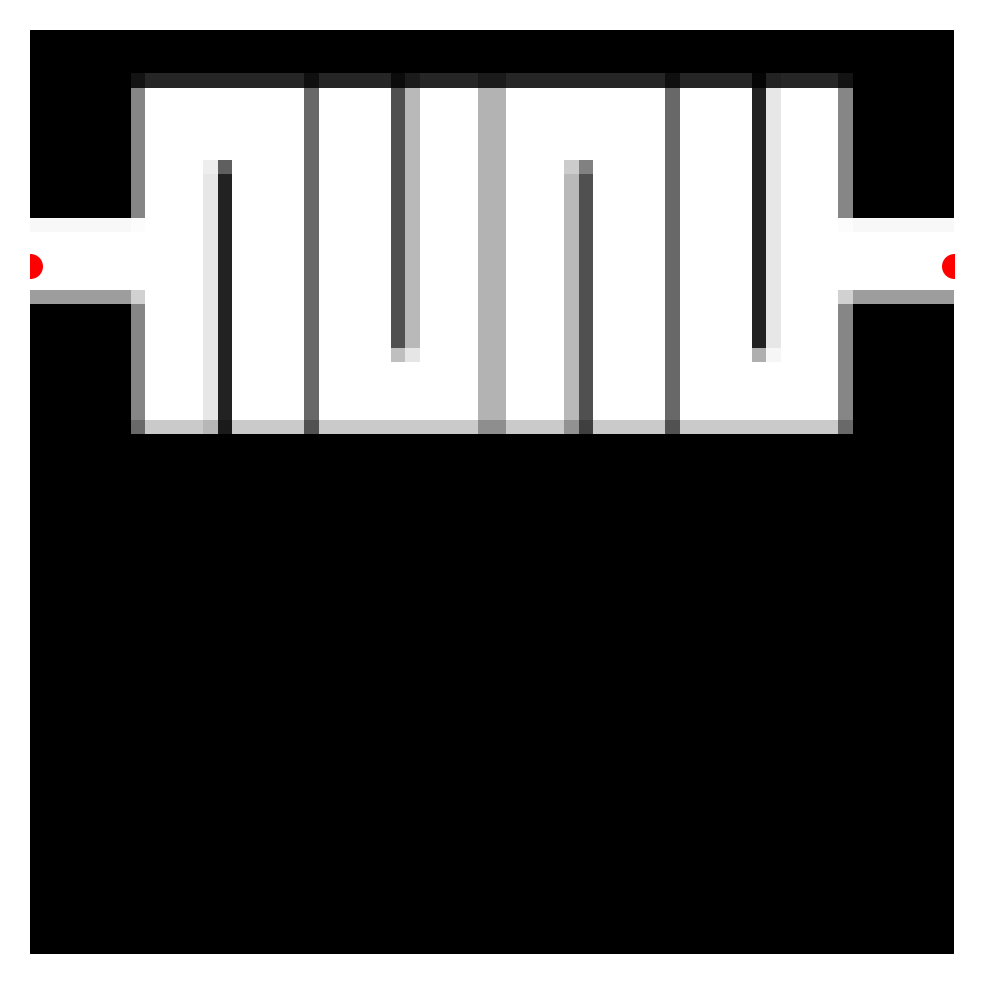}} &
      \raisebox{.2\height}{\includegraphics[width=.2\linewidth]{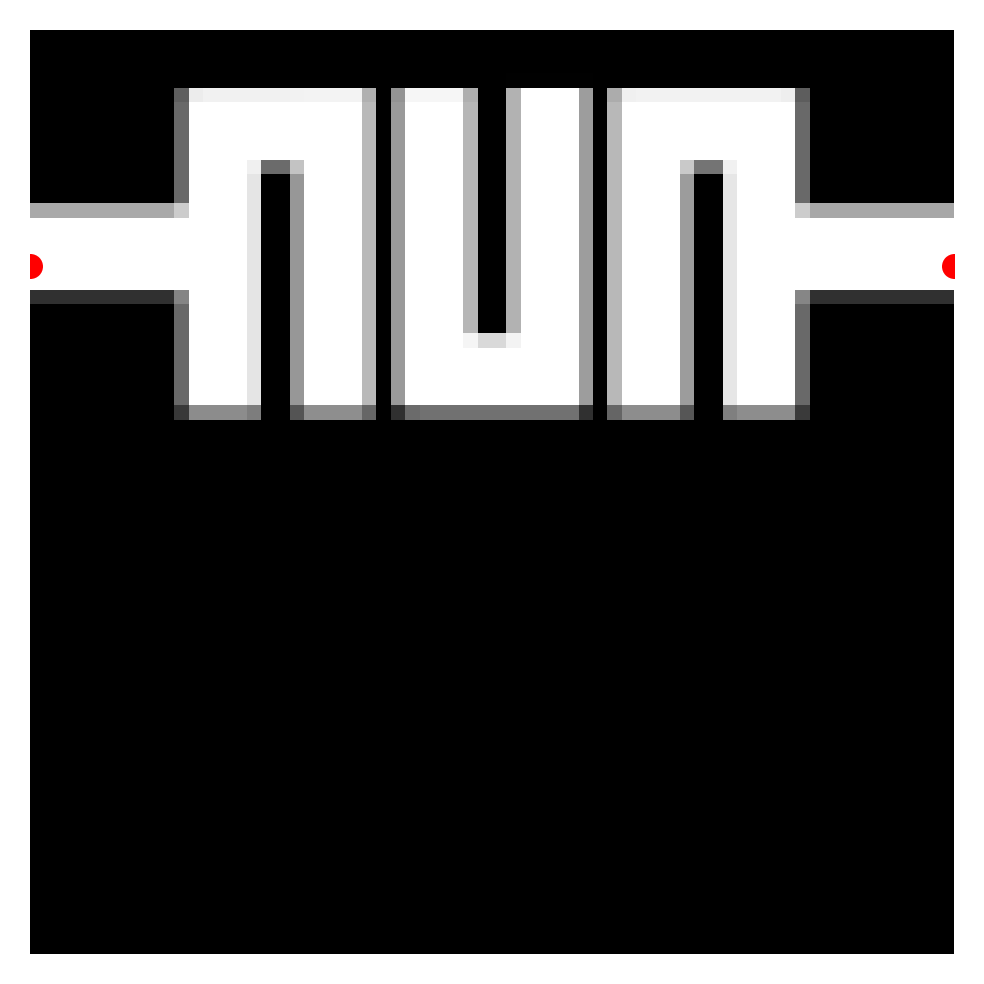}} &
      \raisebox{.2\height}{\includegraphics[width=.2\linewidth]{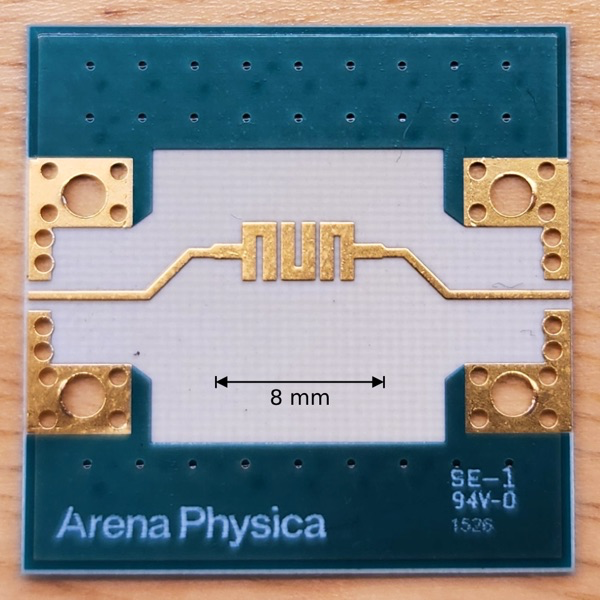}} &
      \raisebox{.1\height}{\includegraphics[width=.35\linewidth]{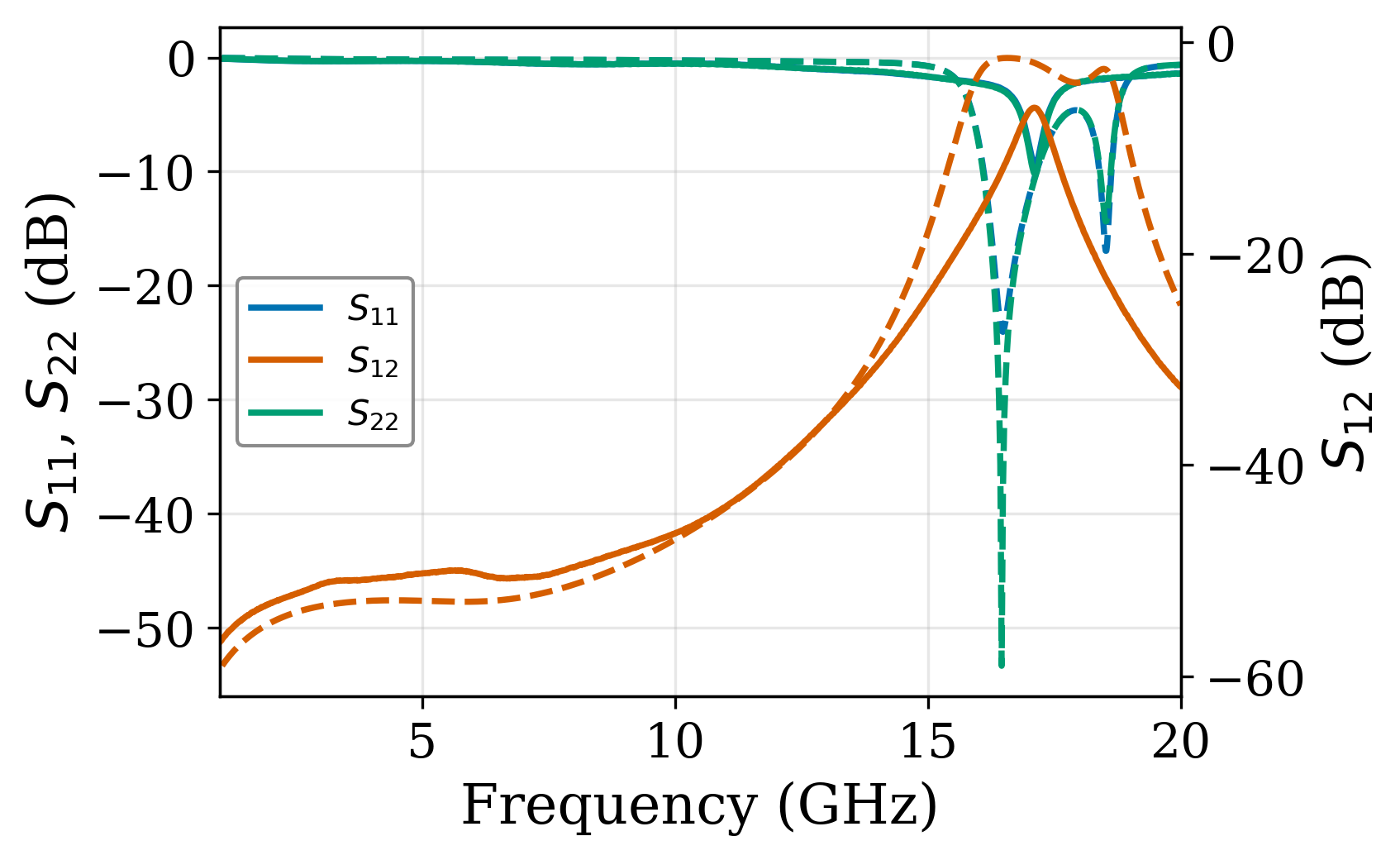}} \\ [-8pt]
      \raisebox{5\height}{N/A} &
      \raisebox{.2\height}{\includegraphics[width=.2\linewidth]{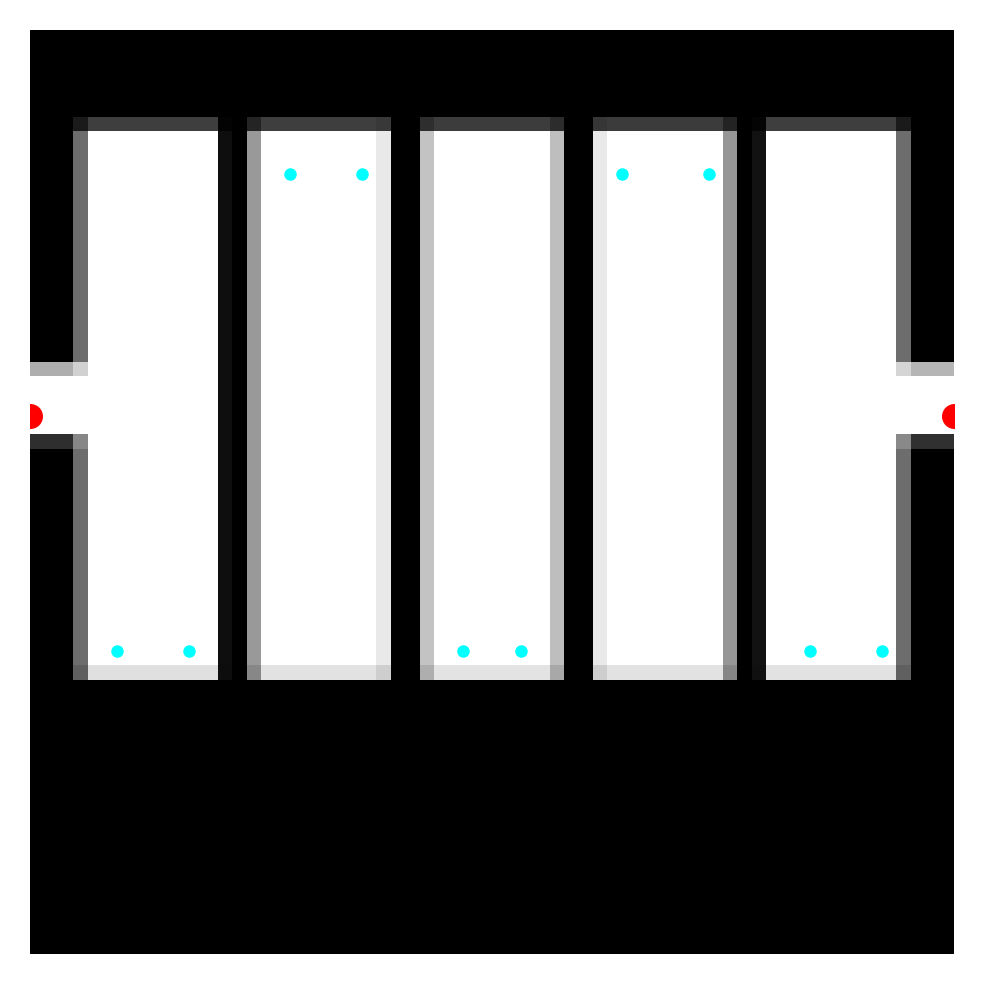}} &
      \raisebox{.2\height}{\includegraphics[width=.2\linewidth]{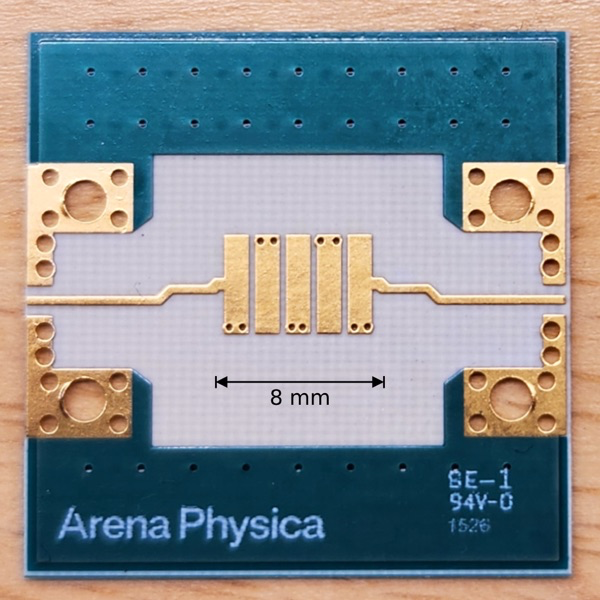}} &
      \raisebox{.1\height}{\includegraphics[width=.35\linewidth]{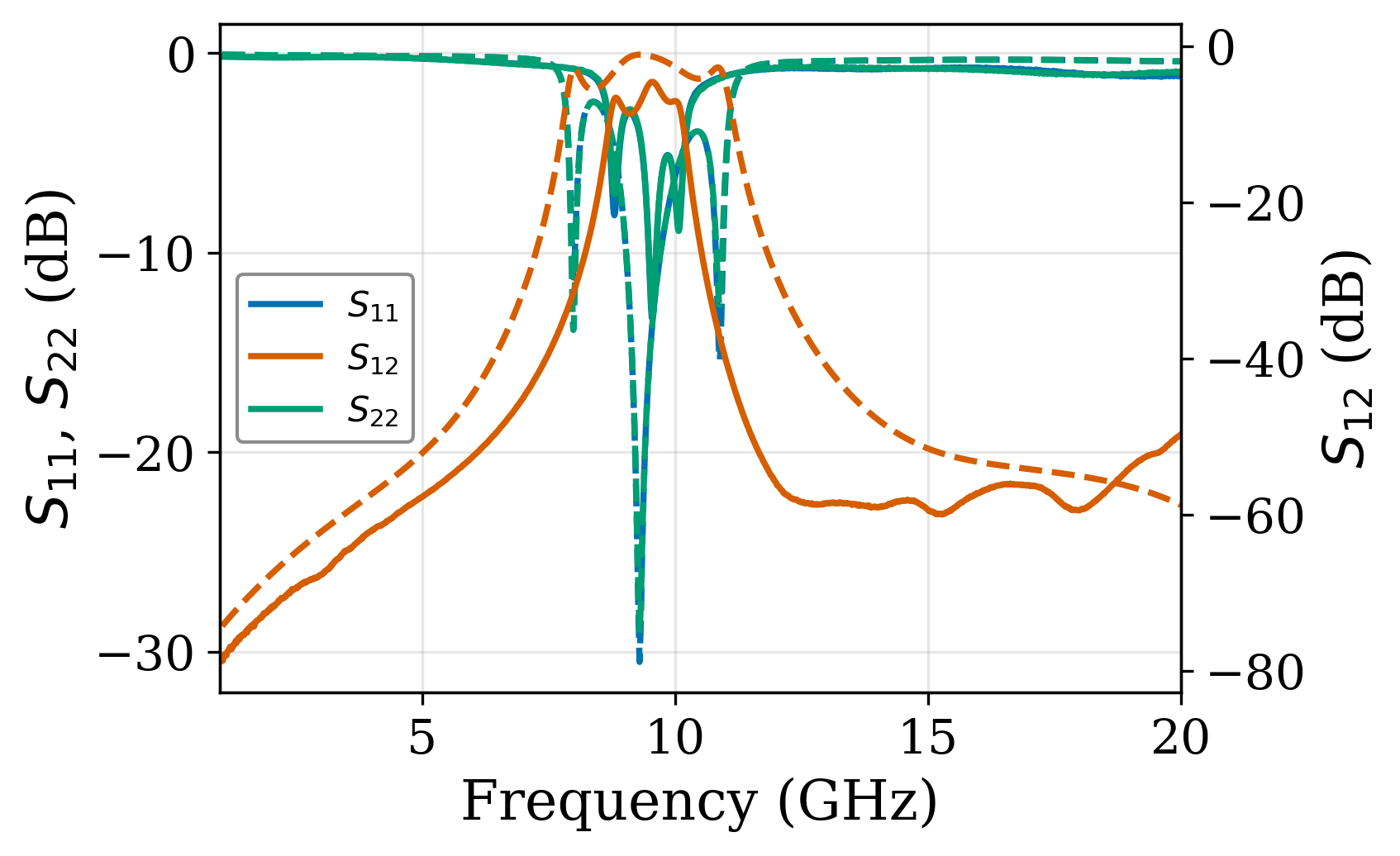}} \\ [-4pt]
    \end{tabular}

    \caption{
    Top: a hairpin filter whose original design has coupling gaps below fabrication limits; the framework generates a manufacturable alternative. Bottom: a combline filter with vias designed from scratch given only target S-parameters (N/A original design). Fabricated boards on RO4003C and
    measured (solid) vs. target (dashed) S-parameter magnitude.}
    \label{fig:hairpin}
\end{figure}

Rather than manually redesigning a hairpin bandpass filter whose coupling gaps fall below low-cost PCB fabrication
limits, we extract its S-parameters and generate manufacturable alternatives. The framework produces multiple
structurally distinct candidates with comparable S-parameter performance (Fig.~\ref{fig:hairpin}, top row), exploiting
the many-to-one nature of the inverse problem.

The manufacturable alternative was fabricated on RO4003C substrate and measured to validate the end-to-end pipeline.
Measured results show a passband centered at approximately 17 GHz---over two guided wavelengths at the board scale---with
6.2 dB minimum insertion loss and 5--9 dB return loss, compared to the target of 1.5 dB insertion loss centered at 17.3
GHz. The center frequency agrees to within 2\%, and the overall filter response shows good agreement with full-wave simulation.

\subsection{Ground-Up Design: Combline Bandpass Filter}

To demonstrate inverse design from scratch, we specify target S-parameters for a combline bandpass filter without
providing any reference layout. Given only the target frequency response, feed locations, and substrate properties, the framework generates candidate designs including via-connected ground structures in seconds (Fig.~\ref{fig:hairpin}, bottom row), exercising the via-inclusive generation capability. The design was fabricated on RO4003C and measured. Measured results show a passband centered at approximately 9.5 GHz---above one guided wavelength at the board scale---with 4.6 dB minimum insertion loss and 10--13 dB return loss, compared to the target of 1.1 dB insertion loss centered at 9.4  GHz. The center frequency agrees to within 1\%, and the overall filter response shows good agreement with full-wave simulation.

\subsection{Comparison with State of the Art}

{
\setlength{\tabcolsep}{8pt}%
\renewcommand{\arraystretch}{1.15}%
\begin{table}[t]
\caption{Comparison with prior inverse EM design methods.}
\centering
\begin{tabular}{l|c|c|c|c|c}
\hline
\textbf{Feature} & \textbf{\cite{guo2025}} & \textbf{\cite{karahan2024}} & \textbf{\cite{zhang2024}} & \textbf{\cite{chenna2025}} & \textbf{Ours} \\
\hline
\multicolumn{6}{l}{\scriptsize\textit{Architecture}} \\
Method       & Diffusion & CNN+GA  & cGAN  & DBS   & \textbf{Diffusion}  \\
Grid         & 18$^2$    & 10--25$^2$ & 32$^2$ & 17$^2$$\times$5 & \textbf{64$^2$} \\
Layers       & 1         & 1       & 1     & Multi & \textbf{Multi}  \\
\hline
\multicolumn{6}{l}{\scriptsize\textit{Design capabilities}} \\
Vias         & No        & No      & No    & Yes*  & \textbf{Yes}    \\
Grayscale    & No        & No      & No    & No    & \textbf{Yes}    \\
Size ($\lambda$) & ${\sim}$0.4 & ${\sim}$0.3 & ${\sim}$0.8 & ${\sim}$0.04 & \textbf{${\sim}$1.0} \\
Hard constraints & No    & No      & No    & N/A   & \textbf{Yes}    \\
\hline
\multicolumn{6}{l}{\scriptsize\textit{Conditioning}} \\
Port cond.   & Edge      & Grid    & Fixed & Fixed & \textbf{Arbitrary}  \\
Template cond. & No      & No      & No    & No    & \textbf{Yes}    \\
Dielectric cond. & No    & No      & No    & No    & \textbf{Yes}    \\
Partial S cond. & No     & No      & No    & N/A   & \textbf{Yes}    \\
\hline
\multicolumn{6}{l}{\scriptsize\textit{Training \& inference}} \\
Training data & 145K      & N/R     & N/R     & N/A   & \textbf{3.5M}   \\
Gen.\ time   & Seconds   & Minutes & Minutes & Hours & \textbf{Seconds} \\
\hline
\multicolumn{6}{l}{\scriptsize *On-chip multi-layer via optimization, not generative ML.}\\
\multicolumn{6}{l}{\scriptsize Electrical size at max operating frequency on native substrate.}\\
\multicolumn{6}{l}{\scriptsize N/R = not reported in publication.}
\end{tabular}
\label{tab:comparison}
\end{table}
}

Table~\ref{tab:comparison} summarizes the comparison. The proposed framework is the only system supporting vias, grayscale metallization, hard physical constraints, and partial S-parameter conditioning simultaneously, on a board approximately one guided wavelength at 20\,GHz ($\lambda_g \approx 8.0$\,mm for $\varepsilon_r = 3.55$), 2.5$\times$ larger than
DALL-EM~\cite{guo2025}. Generation takes approximately 3 seconds (128 candidates, DPM++) or 10 seconds (16 candidates, Langevin) on a single NVIDIA A10G GPU.

\section{Conclusion}

This paper presented a conditional diffusion framework for inverse design of PCB-scale RF passive components that, for
the first time, supports via-inclusive multi-layer designs, grayscale continuous metallization, hard physical
constraints via Langevin projection, and partial S-parameter conditioning. The framework generates diverse candidate
designs in seconds across 25 component templates spanning 1--20 GHz on electrically large boards. Two fabricated designs
on RO4003C validate the end-to-end pipeline: a manufacturable alternative to a hairpin filter with sub-resolution
coupling gaps achieves 6.2 dB insertion loss at 17 GHz, and a combline bandpass filter with vias designed from scratch
given only target S-parameters achieves 4.6 dB insertion loss at 9.5 GHz, both showing good agreement between full-wave
simulation and measurement. The 8$\times$8 mm board with 0.125 mm pixel pitch and grayscale metallization enables sub-pixel
feature resolution critical for modeling tightly coupled structures whose spacing falls below standard fabrication
limits, bridging the gap between generative design and manufacturability.

\end{document}